%% file: ijcai26.tex
\documentclass{article}
\usepackage{ijcai26}

\usepackage{times}
\usepackage{soul}
\usepackage{url}
\usepackage[hidelinks]{hyperref}
\usepackage[utf8]{inputenc}
\usepackage[T1]{fontenc}
\usepackage[small]{caption}
\usepackage{graphicx}
\usepackage{amsmath}
\usepackage{amsthm}
\usepackage{booktabs}
\usepackage{algorithm}
\usepackage{algorithmic}
\usepackage{multirow}
\usepackage[switch]{lineno}
\usepackage{amsmath,amssymb}
\usepackage[normalem]{ulem}
\useunder{\uline}{\ul}{}
\usepackage{tabularx} 
\usepackage{supertabular}
\usepackage{xltabular}
\usepackage{float}
\usepackage{soul}
\usepackage[table,xcdraw]{xcolor}

\title{Emo-LiPO: Listwise Preference Optimization for \\Fine-Grained Emotion Intensity Control in LLM-based Text-to-Speech}

\author{
Yihang Lin$^{1}$\thanks{Equal contribution.}\and
Li Zhou$^{1*}$\thanks{Corresponding author.}\and
Congwei Cao$^{1,5}$\and
Dongchu Xie$^1$\and\\
Xiaoxue Gao$^2$\and
Chen Zhang$^3$\And
Haizhou Li$^{1,3,4,5\dagger}$
\affiliations
$^1$The Chinese University of Hong Kong, Shenzhen\\
$^2$Agency for Science, Technology and Research\\
$^3$National University of Singapore\\
$^4$Shenzhen Research Institute of Big Data\\
$^5$Shenzhen Loop Area Institute\\
\emails
yihanglin@link.cuhk.edu.cn,
\{lizhou21,haizhouli\}@cuhk.edu.cn
}

\begin{document}

\maketitle

\input{section/00-abstract}
\input{section/01-introduction}

\input{section/02-related-work}

\input{section/03-method}
\input{section/04-dataset}
\input{section/05-experiments}

\input{section/06-conclusion}

\section*{Ethical Statement}

There are no ethical issues.

\section*{Acknowledgments}

This research is supported by National Natural Science Foundation of China (Grant No. 62271432), the Program for Guangdong Introducing Innovative and Entrepreneurial Teams(Grant No. 2023ZT10X044), Shenzhen Science and Technology Program(Shenzhen Key Laboratory, Grant No. ZDSYS20230626091302006), and the Shenzhen Stability Science Program 2023. This research was also supported by the Internal Project of the Shenzhen Research Institute of Big Data (SRIBD), the internal project of the Guangdong Provincial Key Laboratory of Big Data Computing under Grant B10120210117-KP02 (The Chinese University of Hong Kong, Shenzhen), and Shenzhen Key Lab of Multi-Modal Cognitive Computing.



\bibliographystyle{named}
\bibliography{ijcai26}

\end{document}

%% file: section/00-abstract.tex
\begin{abstract}

Large language model (LLM)-based text-to-speech (TTS) systems enable prompt-conditioned emotional control but struggle with fine-grained emotion intensity due to the semantic--acoustic gap between text and speech.
To address this challenge, we formulate emotion intensity control in LLM-based TTS as a learning-to-rank problem and propose \textbf{Emo-LiPO}, a listwise preference optimization framework that aligns prompt-conditioned speech generation with relative emotion intensity expressed in text.
Emo-LiPO explicitly models global intensity ordering within each emotion under fixed transcripts, enabling more faithful and continuous emotional expression.
We further construct \textbf{ESD-plus}, a multi-speaker dataset with explicit emotion intensity variations, to support fine-grained emotion modeling and evaluation.
Experiments on ESD-plus demonstrate that Emo-LiPO significantly improves emotion accuracy and intensity controllability over both supervised- and DPO-based LLM TTS baselines, with particularly pronounced gains at high intensity levels.
\end{abstract}

%% file: section/01-introduction.tex
\section{Introduction}


Large language model (LLM)-based text-to-speech (TTS) systems have recently enabled prompt-conditioned control over speaking style and emotion~\cite{leng2023prompttts,du2024cosyvoice,du2024cosyvoice2,zhang2025proemo,yang2025emovoice,11464460}, offering substantially greater flexibility than traditional signal-driven approaches~\cite{zhou2022emotion,wang2023fine,schnell2021improving,im2022emoq,lei2022msemotts,cho2024emosphere,cho2025emosphere++}. Despite this progress, fine-grained control over emotion intensity remains a fundamental challenge for prompt-conditioned LLM-based TTS systems~\cite{guo2023prompttts,leng2023prompttts,du2024cosyvoice,yang2025emovoice}.
\begin{figure}[t]
    \centering
    \includegraphics[width=1.\linewidth,
    trim=0.7cm 2.2cm 2.8cm 10cm,
        clip]{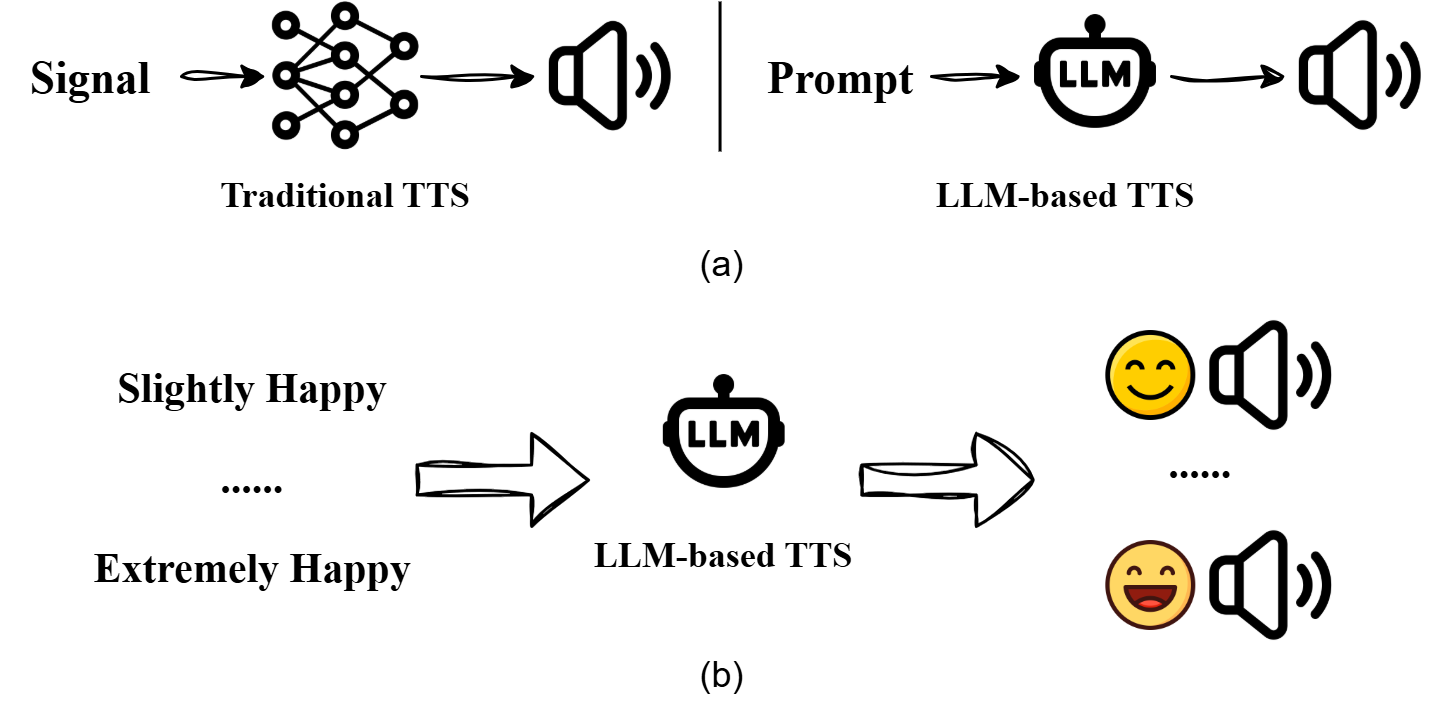}
    \caption{Illustration of the task of prompt-conditioned emotion intensity control in LLM-based TTS.}
    \label{fig:introduction}
\end{figure}

A key difficulty lies in the semantic--acoustic gap between textual emotion descriptions and their realization in speech.
While natural language prompts can effectively convey emotion categories (e.g., \emph{happy} or \emph{sad}), they often fail to reliably induce perceptually distinguishable variations in emotion strength~\cite{guo2023prompttts,leng2023prompttts,zhang2025proemo}, as illustrated in Figure~\ref{fig:introduction}.
As a result, existing models tend to realize emotion intensity implicitly at the acoustic level, leading to unstable or weakly differentiated intensity expressions, especially under high-intensity emotional prompts~\cite{du2024cosyvoice,du2024cosyvoice2,yang2025emovoice}.
This limitation significantly constrains the controllability and expressiveness of emotional TTS systems in real-world applications.

Meanwhile, recent advances in preference optimization have demonstrated strong potential for aligning generative models with subjective human judgments.
Pair-wise and list-wise preference learning frameworks provide principled mechanisms for modeling relative comparisons, making them particularly suitable for perceptual generation tasks~\cite{rafailov2023direct,zhao2023slic,liu2025lipo}.
In the context of TTS, emerging preference-aligned methods have shown improved emotional expressiveness and perceptual quality by encouraging preferred outputs over less preferred ones~\cite{zhang2024speechalign,gao2025emo,liu2025direct,yang2025rlaif,li2025emorl}.
However, existing preference-based TTS approaches primarily focus on controlling emotion at the categorical level, without explicitly modeling the ordinal structure of emotion intensity.
Consequently, relative intensity information specified in text prompts is not directly aligned with preference supervision during training.

In this work, we address the challenge of fine-grained emotion intensity control in LLM-based TTS by framing it as a learning-to-rank problem.
We adopt a listwise preference optimization approach that explicitly aligns prompt-conditioned speech generation with relative emotion intensity conveyed by natural language.
To facilitate systematic modeling and evaluation of emotion intensity, we additionally construct a multi-speaker dataset with explicit intensity variations and conduct extensive experiments to validate the effectiveness of our approach.
Our main contributions can be summarized as follows:
\begin{itemize}
    \item We propose Emo-LiPO\footnote{Code is available at \href{https://github.com/hlt-cuhksz/Emo-LiPO}{hlt-cuhksz/Emo-LiPO}.}, a listwise preference optimization framework that formulates emotion intensity control in LLM-based TTS as a learning-to-rank problem, enabling fine-grained and text-grounded intensity modeling.
    \item We construct ESD-plus\footnote{Dataset is available at \href{https://huggingface.co/datasets/hlt-cuhksz/ESD-plus}{hlt-cuhksz/ESD-plus}.}, a multi-speaker emotional speech dataset with explicit emotion intensity variations to support fine-grained emotion intensity modeling and evaluation.
    \item Extensive experiments on ESD-plus demonstrate that EMO-LiPO outperforms supervised- and DPO-based baselines in emotion accuracy and intensity controllability, with especially pronounced gains at high intensity.
\end{itemize}

%% file: section/02-related-work.tex
\section{Related Work}

\subsection{Emotion Intensity Control in TTS}

Early studies on emotion intensity control mainly operate in the acoustic domain, where emotion strength is explicitly modeled as a continuous variable derived from speech signals using relative ranking or learning-to-rank objectives~\cite{zhou2022emotion,wang2023fine}.
Other approaches extract intensity representations at different temporal levels and inject them into TTS systems for fine-grained control~\cite{schnell2021improving,im2022emoq,lei2022msemotts}.
More recent methods further disentangle emotion intensity from style using geometric representations or incorporate soft intensity guidance through diffusion-based models~\cite{cho2024emosphere,cho2025emosphere++,liang2025ece,guo2023emodiff,wu2024laugh}.
While effective in controlled settings, these approaches typically rely on explicit acoustic features or specialized architectures and lack flexible, text-driven intensity control.
In contrast, recent prompt-conditioned and LLM-based TTS systems leverage natural language descriptions for expressive speech generation, but primarily control discrete emotion categories or coarse prosodic attributes, leaving emotion intensity implicitly realized~\cite{guo2023prompttts,leng2023prompttts,zhang2025proemo,du2024cosyvoice,du2024cosyvoice2,yang2025emovoice}.
Our approach directly grounds emotion intensity in natural language and aligns it with preference supervision, enabling explicit and fine-grained intensity control in LLM-based TTS.


\subsection{Preference Optimization for Generation Tasks}
Preference optimization aims to align pre-trained large language models (LLMs) with human preferences through gradient-based optimization during training. Existing approaches can be broadly categorized into point-wise, pair-wise, and list-wise methods. Point-wise methods optimize models using feedback defined on individual samples, typically in the form of scalar rewards or binary desirability signals, as exemplified by Proximal Policy Optimization (PPO)~\cite{schulman2017proximal}, Reward Maximization (ReMax)~\cite{li2024remax}, and Kahneman–Tversky Optimization (KTO)~\cite{ethayarajh2024kto}. Pair-wise methods, such as Direct Preference Optimization (DPO)~\cite{rafailov2023direct,Zhang-etal-2025-alignfollow} and SLiC-HF~\cite{zhao2023slic}, learn from relative preferences between two candidate outputs by contrasting a preferred response against a less preferred one, shaping the model’s generation distribution accordingly~\cite{zhang-etal-2024-ts}. 
List-wise methods further generalize preference alignment as a ranking problem, directly optimizing list-structured preference signals within a learning-to-rank framework \cite{song2024preference,liu2025lipo}.
These methods motivate us by providing a principled approach to learning from subjective human judgments in natural language processing, making them especially well suited for perceptual generation tasks such as emotional speech synthesis.

\subsection{Preference Alignment for TTS}

Recent studies have begun to incorporate preference alignment into TTS systems to better match human perceptual judgments of speech quality, expressiveness, and emotional appropriateness. SpeechAlign~\cite{zhang2024speechalign} is an early attempt in this direction, aligning codec-based speech language models by learning preferences between golden and synthetic codec tokens to mitigate distribution mismatches between training and inference.
Building upon this paradigm, several works apply preference optimization to emotional speech synthesis. Emo-DPO~\cite{gao2025emo} and ARDM-DPO~\cite{liu2025direct} adopt pair-wise preference objectives to encourage emotionally preferred outputs over less preferred ones, while RLAIF-SPA~\cite{yang2025rlaif} and EMORL-TTS~\cite{li2025emorl} employ reinforcement learning with task-specific rewards to improve emotional expressiveness and controllability. These approaches demonstrate the effectiveness of preference-based optimization for enhancing perceptual speech quality.
However, existing preference-aligned TTS methods mainly focus on selecting better samples rather than explicitly modeling the ordinal structure of emotion intensity. As a result, emotion strength is typically optimized implicitly at the acoustic level, and relative intensity specified in text prompts is not directly aligned with preference signals.

%% file: section/03-method.tex
\section{Method}

\begin{figure*}[htbp]
    \centering
    \includegraphics[
        width=1\linewidth,
        trim=0.7cm 1.05cm 3.7cm 0cm,
        clip
    ]{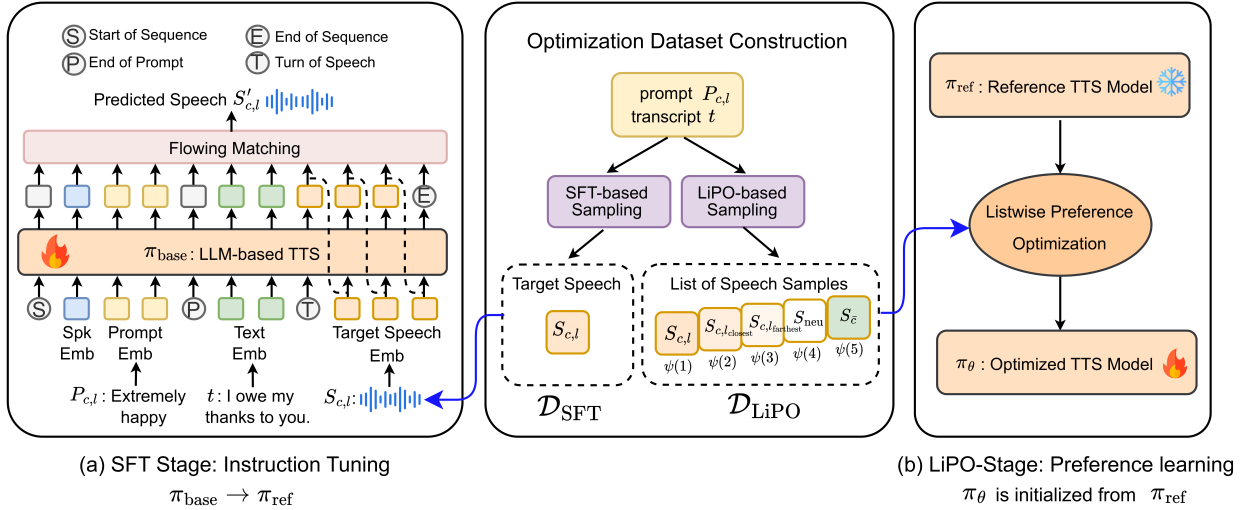}
    \caption{Emo-LiPO framework.
The LLM-based model is trained with SFT followed by LiPO for fine-grained emotion intensity control.}
    \label{fig:method}
\end{figure*}

\subsection{Problem Formulation}

\paragraph{Task Definition.}

We investigate fine-grained control of emotion intensity in LLM-based TTS systems. Given a text transcript and an emotion specification, the TTS model generates speech conditioned on both linguistic content and emotional attributes. Natural language prompts specify an emotion category and, for non-neutral emotions, a relative intensity level. Our goal is to train an LLM-based TTS system to generate speech that accurately follows these specifications.

Formally, let $\mathcal{C} = \mathcal{C}_{\text{emo}} \cup \{\text{\textit{neutral}}\}$ denote the set of emotion categories, where $\mathcal{C}_{\text{emo}} = \{\text{\textit{happy}}, \text{\textit{sad}}, \text{\textit{angry}}, \ldots\}$ and \textit{neutral} corresponds to emotionally flat speech. For each $c \in \mathcal{C}_{\text{emo}}$, we define an ordered set of intensity levels $\mathcal{L} = \{l_1, l_2, \ldots, l_K\}$, where $l_i < l_j$ for $i < j$ indicates strictly increasing emotional intensity (e.g., slightly, moderately, extremely). Crucially, the weakest level $l_1$ must remain perceptually distinct from neutral speech, preserving a clear boundary between emotional and non-emotional outputs.

We define the model input as $x = (t, P_{c,l})$, consisting of a text transcript $t$ and an emotion prompt $P_{c,l}$ specifying an emotion category $c$ and intensity level $l$.
The TTS model generates speech $S = \pi_\theta(x)$, which should satisfy:
(i) \textbf{Content fidelity}, meaning the speech faithfully conveys the linguistic content of the transcript $t$;
(ii) \textbf{Category correctness},  
meaning the generated speech follows emotion $c$ exactly; and
(iii) \textbf{Intensity ordering} for $c \in \mathcal{C}_{\mathrm{emo}}$, such that speech generated with higher intensity levels is perceived as strictly stronger than that generated with lower levels.

\paragraph{Learning-to-Rank Formulation via LiPO (Emo-LiPO).}
\label{sec:lipo}

To meet these requirements, we adopt the Listwise Preference Optimization (LiPO) framework~\cite{liu2025lipo} and formulate fine-grained emotion intensity control as a Learning-to-Rank (LTR) problem. 
In this formulation, all candidate speech samples compared under the same prompt are generated from the same text transcript.
LiPO learns from listwise preference data defined over multiple candidates generated for the same prompt, making it well suited for modeling both emotion category correctness and global intensity ordering.

In contrast, Supervised Fine-Tuning (SFT) relies on independent prompt–target pairs, $\mathcal{D}_{\mathrm{SFT}}=\left\{\left( x = (t, P_{c,l}),\, y = S_{c,l} \right)\right\}$, which provide no supervision on relative emotional intensity across samples. Direct Preference Optimization (DPO) extends SFT by introducing pairwise preferences, $\mathcal{D}_{\mathrm{DPO}}=\left\{
\left( x = (t, P_{c,l}),\, y = S_{c,l}, \bar{S}_{c,l}) \right)
\right\}$,
where $\bar{S}_{c,l}$ denotes a negative sample. While DPO captures local preference information, it remains limited to binary comparisons. In emotional TTS, existing DPO-based methods such as Emo-DPO~\cite{gao2025emo} similarly rely on pairwise emotional preferences\footnote{In Emo-DPO, each training instance is defined as $x=(t, P_c)$, $y=(S_c, S_{\bar{c}})$.}. As a result, these approaches are inherently constrained to local comparisons and cannot enforce a global ordering of emotional intensity.

Notably, Emo-LiPO adopts a listwise formulation, $\mathcal{D}_{\mathrm{LiPO}}=\left\{\left( x = (t, P_{c,l}),\, y = (\mathcal{T}_{c,l}, \psi_{c,l}) \right)\right\}$,
where $\mathcal{T}_{c,l}$ is a list of speech samples generated by the TTS model, and
$\psi_{c,l} \in [0,1]^{|\mathcal{T}_{c,l}|}$ is a vector of real-valued preference scores associated with the corresponding samples, with larger values indicating stronger preference.
This listwise supervision enables Emo-LiPO to explicitly model fine-grained emotional intensity rankings across multiple candidates.

\subsection{Emo-LiPO Framework}

We present the Emo-LiPO framework by specifying the construction of listwise preference data and the corresponding optimization objective under the LiPO formulation.

\paragraph{Rule-Based Preference Construction.}

For a prompt $P_{c,l}$, we construct a listwise preference set $\mathcal{T}_{c,l}$ of $K+2$ speech candidates with the same text transcript $t$ using a \textbf{rule-based ranking strategy}. The list consists of:
(i) a \textit{target} sample $S_{c,l}$ with the exact emotion category $c$ and intensity level $l$;
(ii) $K-1$ \textit{same-emotion} samples with the same category $c$ but different intensity levels $l' \in \mathcal{L} \setminus \{l\}$;
(iii) one \textit{neutral} speech sample; and
(iv) one \textit{negative} sample with a randomly selected non-target emotion category.
Under this rule-based strategy, the $K-1$ \textit{same-emotion} samples are ordered according to their absolute intensity distance $|l' - l|$, with samples closer to the target intensity being preferred; ties are resolved by random ordering. The resulting preference list follows the order
\begin{equation}
\mathcal{T}_{c,l} =
\left[
S_{c,l}
\succ S_{c,l_{\text{closest}}}
\succ \cdots
\succ S_{c,l_{\text{farthest}}}
\succ S_{\mathrm{neu}}
\succ S_{\bar{c}}
\right],
\label{eq:T-cl}
\end{equation}
where $l_{\text{closest}}$ and $l_{\text{farthest}}$ denote intensity levels of the same emotion category ordered by increasing distance from $l$.

Based on the rule-based ranking described above, we assign a real-valued preference label vector
$\psi_{c,l} \in [0,1]^{|\mathcal{T}_{c,l}|}$ to the speech samples in $\mathcal{T}_{c,l}$.
Specifically, we adopt an \textbf{index-based} preference scoring scheme, where the preference label of each sample is determined by its position in the ordered list $\mathcal{T}_{c,l}$.
Let $i$ denote the index of a sample in the list, with smaller indices indicating stronger preference.
The preference label of the $i$-th sample is defined as
\begin{equation}
\psi_{c,l}(i) = 1 - \frac{i-1}{K+2},
\quad i = 1, \ldots, K+2.
\label{eq:psi-cl}
\end{equation}

\paragraph{Multi-Stage Optimization.}
As illustrated in Figure~\ref{fig:method}, following Emo-DPO~\cite{gao2025emo}, we optimize the TTS model in a multi-stage manner, starting with SFT to learn basic instruction following, and subsequently applying LiPO for fine-grained emotion intensity control.
The LLM-based TTS model formulates text-to-speech as a conditional autoregressive generation task~\cite{du2024cosyvoice}, in which speech tokens are generated sequentially conditioned on the joint input
$x = (t, P_{c,l})$, together with a speaker embedding.

Specifically, in the SFT stage, we perform supervised fine-tuning on a paired prompt--speech dataset $\mathcal{D}_{\mathrm{SFT}}$ to initialize the backbone TTS model $\pi _{\mathrm{base}}$:
\begin{equation}
    \mathcal{L} _{\mathrm{SFT}}(\pi_{\mathrm{base}} )=\mathbb{E} _{(x,S)\sim \mathcal{D} _{\mathrm{SFT}}}\left[ -\log \pi _{\mathrm{base}}(S\mid x) \right], 
\end{equation}
where $\log \pi_{\mathrm{base}}(S \mid x)$ is computed using teacher forcing with token-level cross-entropy over the autoregressively generated speech tokens.
The resulting model is denoted as the reference TTS policy $\pi_{\mathrm{ref}}$ in the subsequent stage.

In the LiPO stage, initialized from $\pi_{\mathrm{ref}}$, we optimize the model using the listwise preference dataset $\mathcal{D}_{\mathrm{LiPO}}$ as defined in \S\ref{sec:lipo}. Given a preference list $\mathcal{T}_{c,l}$ and its associated preference label vector $\psi_{c,l}$, we apply Listwise Preference Optimization to encourage the model to generate speech samples that better align with the desired emotion intensity ordering. The corresponding optimization objective is defined as:
\begin{equation}
\mathcal{L} _{\mathrm{LiPO}}(\pi _{\theta};\pi _{\mathrm{ref}},\beta )=\mathbb{E} _{(x,\mathcal{T} _{c,l},\psi _{c,l})\sim \mathcal{D} _{\mathrm{LiPO}}}\left[ r\left( \psi _{c,l},\mathrm{s} \right) \right],
\end{equation}
where $r(\cdot)$ denotes a listwise learning-to-rank loss under the LiPO framework, and $s$ represents the ranking scores induced by the current policy $\pi_{\theta}$ relative to the reference policy $\pi_{\mathrm{ref}}$, The score of each candidate $S_i\in \mathcal{T}_{c,l}$ is computed as:
\begin{equation}
s_i=\;\beta \log \frac{\pi _{\theta}\!\left( S_i\mid x \right)}{\pi _{\mathrm{ref}}\!\left( S_i\mid x \right)}.
\label{eq:s_i}
\end{equation}
Following the original LiPO formulation~\cite{liu2025lipo}, $r(\cdot)$ is defined as:
\begin{equation}
r\!\left(\boldsymbol{\psi}_{c,l},\, \mathbf{s} \right)
=
-\sum_{(i,j)\in \mathcal{\psi}_{c,l}}
\lambda_{i,j}\,\big(s_i - s_j\big),
\label{eq:loss}
\end{equation}
where $\mathcal{\psi}_{c,l}=\{(i,j)\mid \psi_{c,l}(i)>\psi_{c,l}(j)\}$, the relative orders of candidate pairs ($S_i$, $S_j$). The weighting term $\lambda_{i,j}$ serves as an intensity-aware coefficient that scales each pairwise loss term according to the implied emotion-intensity discrepancy between $(S_i,S_j)$, injecting fine-grained intensity information beyond the binary preference order and promoting ordinal consistency for more reliable intensity control. We then define the gain and discount functions, $G(\cdot)$ and $D(\cdot)$, based on the position index $i$:
\begin{equation}
G(i)=2^{\psi_{c,l}(i)}-1,\qquad 
D(i)=\frac{1}{\log(1+i)}.
\end{equation}
and $\lambda_{i,j}$ is derived as:
\begin{equation}
\lambda_{i,j}
=
\left|G(i) - G(j)\right|
\cdot
\left|
\frac{1}{D(i)} - \frac{1}{D(j)}
\right|.
\label{eq:lambda}
\end{equation}
A larger $\lambda_{i,j}$ implies a larger implied intensity gap between $S_i$ and $S_j$ under the list ranking, so the model is penalized more strongly for violating their order.

%% file: section/04-dataset.tex
\section{ESD-plus Dataset}


\subsection{Dataset Overview}

To evaluate the effectiveness of the proposed Emo-LiPO method, we construct \textbf{ESD-plus}, a speech dataset with explicit emotion intensity variations designed to support research on fine-grained emotion control in TTS systems.
ESD-plus is built upon the English portion of the ESD corpus~\cite{zhou2022emotional}, from which we select 350 parallel sentences as synthesis scripts.
We retain the original ESD emotion categories, including four non-neutral emotions (happy, surprise, sad, and angry) and one neutral emotion.
For each non-neutral emotion, we further introduce three distinct intensity levels, resulting in a total of 13 fine-grained emotion labels.

In total, ESD-plus contains 45{,}500 fine-grained emotional speech samples, with an average duration of 2.92 seconds per sample, amounting to approximately 36.89 hours of audio for training and evaluation.
Following the official data partitioning strategy of ESD~\cite{zhou2022emotional}, we split the dataset into training, development, and test sets with 300, 20, and 30 utterances, respectively, while ensuring that all speaker–emotion variants of each utterance are assigned to the same split.

\input{table/overall_results}

\subsection{Dataset Construction and Quality Control}

We construct ESD-plus using a prompt-based emotional speech synthesis pipeline built upon \texttt{gpt-4o-mini-tts}\footnote{https://openai.fm/}.
For each emotion--intensity category, we design a structured prompt that combines a natural-language emotion description with a scalar intensity indicator as the conditional control signal. 
To ensure perceptually distinguishable emotional variations across different intensity levels, the scalar indicators are deliberately spaced (e.g., 1/5, 3/5, and 5/5) rather than uniformly incremental.\footnote{Details on dataset construction, quality control, and dataset statistics are provided in the supplementary material (Appendix A).}
Moreover, we generate speech using 10 official English voices, comprising 4 male and 6 female speakers, enabling multi-speaker emotional synthesis.

To ensure data reliability, we perform human verification on the development and test sets of ESD-plus, which are used exclusively for model evaluation.
Annotators perform pairwise comparisons within each emotion category to verify that speech samples generated at different intensity levels exhibit consistent and monotonically increasing emotional intensity in accordance with their prompts.
Samples that fail to demonstrate the expected intensity ordering are flagged and regenerated until the target ordering is satisfied, thereby ensuring the correctness of emotion--intensity labels in the development and test sets.
Overall, 91.23\% of the evaluated samples pass verification, demonstrating that ESD-plus provides high-quality and reliable data for model training and evaluation.\footnote{The training set is not modified during this process, as its overall label correctness is already high.}

%% file: table/overall_results.tex
\begin{table*}[ht]
\centering
\scalebox{1.}{
\begin{tabular}{@{}l|rrrr|rrr@{}}
\toprule
\multirow{2}{*}{\textbf{Model}} & \multicolumn{4}{c|}{\textbf{Speech Quality}} & \multicolumn{3}{c}{\textbf{Emotion Relevance}} \\ \cmidrule(l){2-8}
 & \textbf{WER $\downarrow$} & \textbf{NISQA $\uparrow$} & \textbf{DNSMOS $\uparrow$} & \textbf{UTMOS $\uparrow$} & \textbf{EmoSIM $\uparrow$} & \textbf{Recall $\uparrow$} & \textbf{Recall-ft $\uparrow$} \\ \midrule
\textbf{CosyVoice}      & 4.47  & 4.71          & 3.16          & \textbf{4.30}          & 81.87          & 25.10          & 29.90 \\
\textbf{EmoVoice}      
       & 5.40  & 4.79          & \textbf{3.29} & 4.27          & 89.84          & 20.56          & 28.51 \\ \midrule
\textbf{Emo-DPO (R)}      
       & 12.78  & 4.37          & 3.04          & 3.90          & 91.52          & 24.77          & 33.46 \\
\textbf{Emo-DPO (E)}      
       & 4.78  & 4.66          & 3.23          & 4.06          & 91.73          & 24.08          & 34.92 \\
\textbf{Emo-DPO (I)}  
       & 6.79  & 4.60          & 3.21          & 4.00           & 91.85          & 26.87          & 37.21 \\ \midrule
\textbf{Emo-LiPO}
       & 4.26  & \textbf{4.79}          & 3.26          & 4.18          & 91.93          & \textbf{27.56} & \textbf{39.54} \\
\textbf{\quad\quad - w/o $\lambda$ }      
       & \textbf{4.15}  & 4.79          & 3.24          & 4.17          & \textbf{92.30} & 26.21          & 37.59 \\
\bottomrule
\end{tabular}}
\caption{Comparison of different prompt-conditioned TTS models in terms of speech quality and emotion relevance.}
\label{tab:model_comparison}
\end{table*}

%% file: section/05-experiments.tex
\section{Experiments}

\subsection{Experimental Setup}


\paragraph{Baselines and Implementation.}

We adopt several LLM-based TTS models as baselines.
Specifically, we include two supervised LLM-based TTS models: \textbf{CosyVoice}~\cite{du2024cosyvoice}, which is built upon supervised semantic tokens, and \textbf{EmoVoice}~\cite{yang2025emovoice}, which supports freestyle text prompting for emotional control.
In addition, we consider preference-optimized LLM-based TTS methods following Emo-DPO~\cite{gao2025emo}, which primarily target coarse-grained emotion category control.
To evaluate their effectiveness in modeling fine-grained emotion intensity, we design three Emo-DPO variants:
(i) \textbf{Emo-DPO (R)}, trained with randomly sampled pairwise preferences $( S_{c,l}, \bar{S}_{c,l} )$;
(ii) \textbf{Emo-DPO (E)}, trained using pairwise preferences across different emotion categories at the same intensity level $( S_{c,l}, S_{\bar{c},l} )$;
and (iii) \textbf{Emo-DPO (I)}, trained using pairwise preferences across different intensity levels within the same emotion category $( S_{c,l}, S_{c,\bar{l}} )$.
Our \textbf{Emo-LiPO} model is built upon \textbf{CosyVoice} as the backbone, specifically using the \texttt{CosyVoice-300M-Instruct} version.

\paragraph{Evaluation Metric.}

We evaluate the generated speech from two complementary perspectives: speech quality and emotion relevance.
At the speech level, we assess intelligibility and perceptual quality.
Speech intelligibility is measured using Word Error Rate (\textbf{WER}), computed with the Whisper-Large-v3 ASR model~\cite{radford2023robust}.
Speech naturalness and overall perceptual quality are evaluated using three objective metrics: \textbf{NISQA}~\cite{mittag2021nisqa}, \textbf{DNSMOS}~\cite{reddy2021dnsmos}, and \textbf{UTMOS}~\cite{saeki2022utmos}.
At the emotion level, we evaluate how well the generated speech conveys the intended emotional content.
We employ the emotion2vec model~\cite{ma2024emotion2vec} for speech emotion recognition (SER) and adopt two metrics: \textbf{Emotion Similarity} (\textbf{EmoSIM}) and \textbf{Recall}.
EmoSIM is computed as the cosine similarity between emotion embeddings extracted from the generated and ground-truth speech.
Recall measures the emotion classification accuracy of emotion2vec on the generated speech.
Due to its limited out-of-domain performance, we further fine-tune emotion2vec on ESD-plus and report the resulting score as \textbf{Recall-ft}.

\subsection{ Emotion Category Control Performance}

\paragraph{Effect of Intensity-Aware Preference Modeling.}

Table~\ref{tab:model_comparison} summarizes the performance of different prompt-conditioned TTS models in terms of emotion category control, together with their speech quality metrics.
Overall, Emo-LiPO demonstrates the strongest ability to generate speech that aligns with the intended emotion categories, consistently outperforming both supervised baselines and DPO-based methods, while maintaining comparable speech quality.
Compared with supervised LLM-based TTS models, preference-optimized approaches show clear advantages in emotion relevance, indicating that preference signals are critical for effective emotional control without sacrificing speech naturalness.
Among the DPO-based variants, models trained with structured emotion-related preferences achieve better emotion category recognition than those trained with randomly sampled pairs, highlighting the importance of informative preference construction.
Notably, introducing intensity-aware preference modeling further improves emotion category control.
Emo-DPO trained with intensity-level preferences performs better than variants relying solely on cross-category comparisons, suggesting that modeling relative intensity ordering within emotions provides transferable supervision for category-level emotion recognition.
Finally, Emo-LiPO consistently achieves the best performance across emotion relevance metrics, demonstrating that listwise preference optimization enables more accurate and robust emotion category control under fine-grained emotional prompting, without degrading speech quality.

\paragraph{The impact of weighting term $\lambda$.} 
To analyze the role of the distance-based weighting term $\lambda$ (Eq.~\ref{eq:lambda}) in learning stable intensity ordering, we conduct an ablation study in which $\lambda$ is fixed to 1 for all preference pairs, resulting in a uniformly weighted LiPO loss, while keeping all other training settings unchanged.
As shown in Table~\ref{tab:model_comparison}, adopting uniform weighting leads to a slight improvement in emotion similarity but consistently degrades emotion control metrics compared with the full model.
These findings indicate that $\lambda$  does not primarily improve category-level alignment; rather, it provides more informative supervision for intensity-aware ranking by emphasizing ordinal consistency across candidates, thereby enhancing controllability and generalization during evaluation.

\subsection{Emotion Intensity Control Performance}

\begin{figure}[t]
    \centering
    \includegraphics[width=\columnwidth]{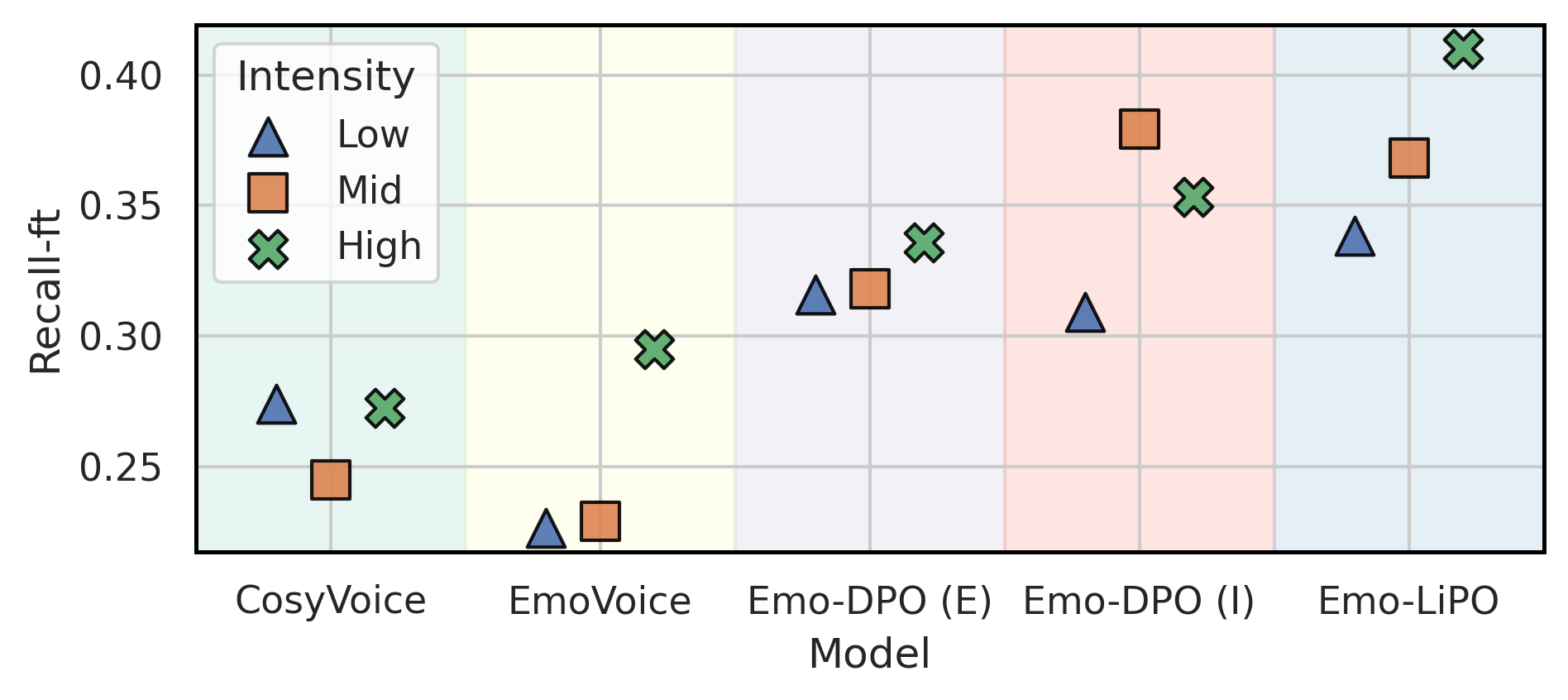}
    \caption{Emotion recognition accuracy (Recall-ft) across different emotion intensity levels.}
    \label{fig:recall_rate}
\end{figure}

\paragraph{Automatic Evaluation Comparison.}

Intuitively, higher emotional intensity leads to more perceptually salient speech and should therefore result in higher emotion recognition accuracy.
Accordingly, we evaluate emotion recognition accuracy (Recall-ft) across different emotion intensity levels, as shown in Figure~\ref{fig:recall_rate}.
Emo-LiPO achieves the strongest overall performance and is the only model that exhibits a clear and stable monotonic increase in Recall-ft from low to high intensity levels.
In contrast, supervised baselines and DPO-based variants display weaker or non-monotonic trends, indicating inconsistent alignment between textual intensity descriptions and the generated speech.
These results demonstrate that Emo-LiPO more faithfully translates relative emotion intensity specifications into perceptually distinguishable speech, validating its effectiveness in fine-grained emotion intensity control.


\paragraph{Human Evaluation Comparison.}
To further validate the effectiveness of Emo-LiPO in emotion intensity control, we conduct a human Arena-style evaluation.
We perform pairwise comparisons between Emo-LiPO and three representative baselines (CosyVoice, Emo-DPO (E), and Emo-DPO (I)) along three perceptual dimensions: \textit{Speech Quality}, \textit{Emotion Expression}, and \textit{Intensity Control}.
For each comparison, annotators are presented with the same text prompt and two candidate models, each providing a pair of speech samples generated under the same script and emotion category at adjacent intensity levels.
The evaluated intensity transitions include
$(S_{\text{neu}}, S_{c,l_{\text{low}}})$, $(S_{c,l_{\text{low}}}, S_{c,l_{\text{mid}}})$, and $(S_{c,l_{\text{mid}}}, S_{c,l_{\text{high}}})$.
Annotators select the model that performs better with respect to a single evaluation dimension, with a tie option allowed when performance is comparable.
The order of models is randomized, and five annotators participate in the evaluation. \footnote{Detailed human evaluation settings, including the evaluation scale, platform, and annotation guidelines, are provided in the supplementary material (Appendix B).}

Table~\ref{tab:arena} reports the Arena win rates of Emo-LiPO compared with three baseline models across three perceptual dimensions.
Overall, Emo-LiPO consistently outperforms all baselines in all evaluation dimensions, demonstrating strong perceptual advantages under human judgment.
When compared with Emo-DPO (I), the advantage of Emo-LiPO in intensity control is less pronounced, suggesting that incorporating intensity-aware pairwise preferences already captures part of the fine-grained intensity information.
However, Emo-LiPO consistently outperforms Emo-DPO (I) in \textit{Speech Quality}, indicating that listwise preference optimization better preserves acoustic naturalness while maintaining competitive intensity controllability.
This comparison highlights a favorable trade-off achieved by Emo-LiPO, where global intensity ordering is modeled without sacrificing speech quality.





\begin{table}[t]
\centering
\resizebox{\columnwidth}{!}{
\begin{tabular}{@{}lrrr@{}}
\toprule
& \textbf{CosyVoice} & \textbf{Emo-DPO (E)} & \textbf{Emo-DPO (I)} \\ 
\midrule
\textbf{Speech Quality}      & 94.29 & 79.37 & 89.28 \\
\textbf{Emotion Expression}  & 90.34 & 80.65 & 78.24 \\
\textbf{Intensity Control}   & 86.08 & 66.44 & 58.33 \\
\bottomrule
\end{tabular}}
\caption{Arena win rate (\%) of Emo-LiPO compared against baseline models across three evaluation dimensions.}
\label{tab:arena}
\end{table}


\begin{figure}[t]
    \centering
    \includegraphics[width=\columnwidth]{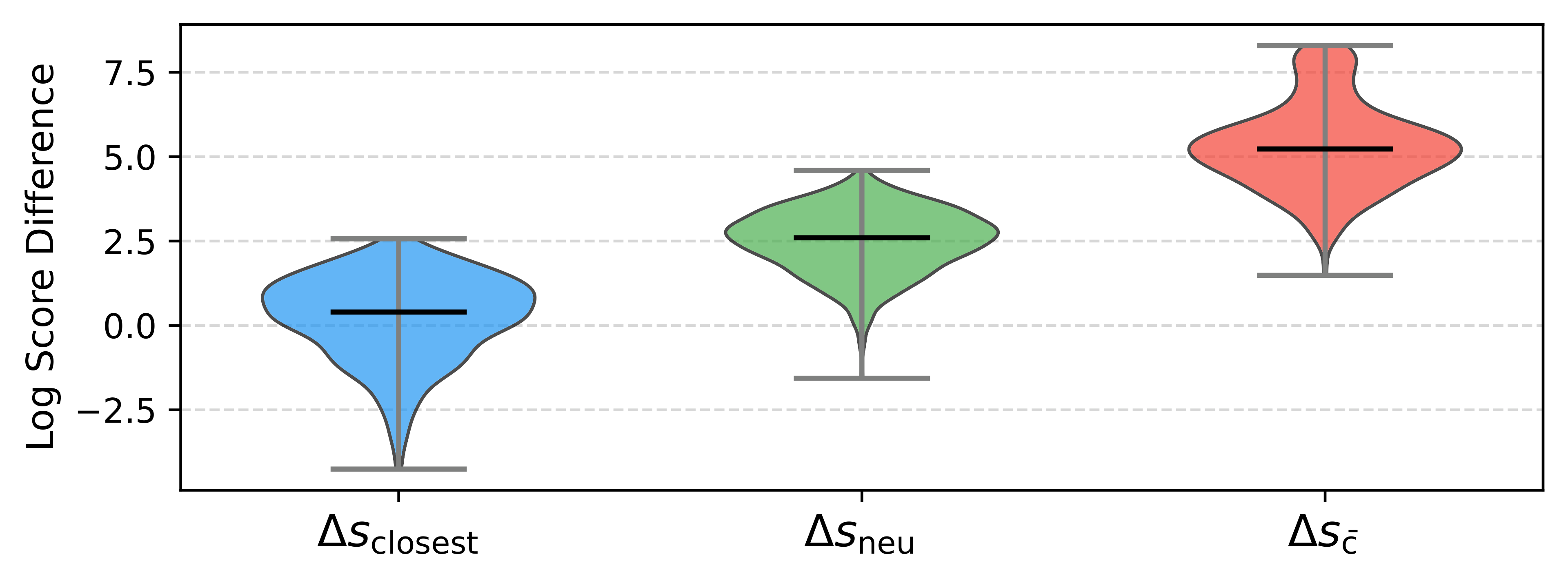}
    \caption{Score Margin Distributions Between Target and Different Candidates.}
    \label{fig:score_distribution}
\end{figure}

\begin{figure*}[ht]
    \centering
    \includegraphics[width=1.\linewidth]{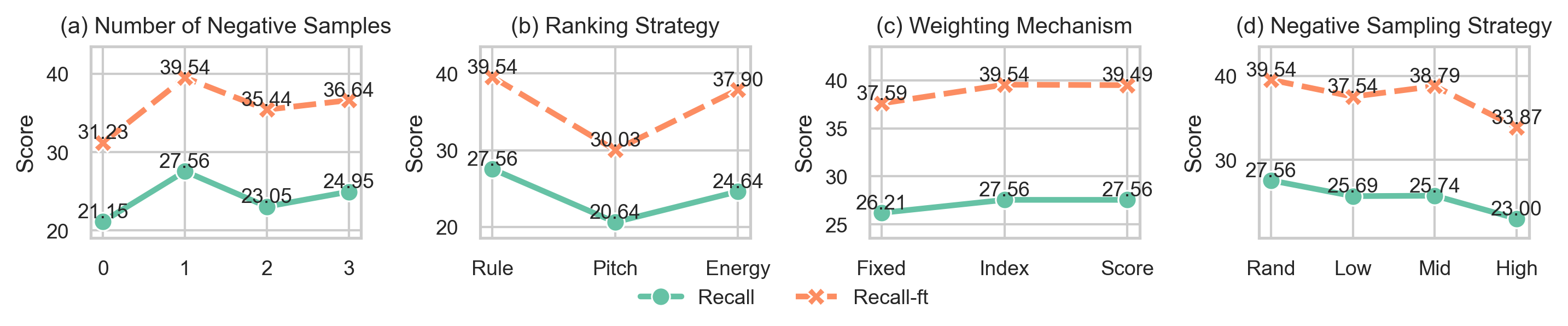}
    \caption{Ablation study of Emo-LiPO on different design choices.}
    \label{fig:analysis}
\end{figure*}
\paragraph{Score Distance Visualization.}
To diagnose whether the Emo-LiPO policy $\pi_{\theta}$ induces a separable preference structure consistent with the rule-based list supervision in Eq.~\ref{eq:T-cl}, we visualize the distribution of score margins between the \textit{target} sample $S_{c,l}$ and the candidates in $\mathcal{T}_{c,l}$. 
Based on Eq.~\ref{eq:s_i}, we compute three absolute margins aligned with the list components:
(i) the margin to closest same emotion sample, $\Delta s_{\text{closest}}=\left|s(S_{c,l})-s(S_{c,l_{\text{closest}}})\right|$;
(ii) the margin to neutral sample, $\Delta s_{\text{neu}}=\left|s(S_{c,l})-s(S_{\mathrm{neu}})\right|$; and
(iii) the margin to other emotion sample, $\Delta s_{{\bar{c}}}=\left|s(S_{c,l})-s(S_{\bar{c}})\right|$.
For comparability across scales, we plot the logarithm of these margins.
As shown in Fig.~\ref{fig:score_distribution}, the three margins exhibit a stable hierarchy in both location and central tendency: $\Delta s_{\bar{c}}$ is the largest and most concentrated, followed by $\Delta s_{\text{neu}}$, while $\Delta s_{\text{closest}}$ is the smallest and closer to zero with a longer tail, reflecting the finer and more challenging discrimination between nearest intensities within the same emotion. 
Overall, this distributional hierarchy is consistent with the rule-based order in Eq.~\ref{eq:T-cl}, indicating that Emo-LiPO shapes a scoring landscape that prioritizes cross-emotion separation while preserving fine-grained intensity hierarchy modeling within the same emotion.


\subsection{Ablation Studies and Analysis}

\paragraph{Effect of the Number of Negative Samples $S_{\bar{c}}$.}
We investigate how the number of negative samples affects intensity learning in listwise optimization by varying the number of negatives while keeping the target intensity chain fixed.
Each negative sample is drawn from a non-target emotion category under the same text condition.
As shown in Figure~\ref{fig:analysis} (a), introducing a single negative sample yields the best overall performance and produces a clearer monotonic trend across intensity levels.
In contrast, increasing the number of negative samples to two or three does not lead to further gains and instead degrades performance.
These results suggest that a limited amount of cross-emotion contrast is sufficient for effective intensity learning, whereas excessive negative samples weaken the supervision signal for modeling intensity ordering.

\paragraph{Effect of Ranking Strategy for Preference List Construction.}

To study how the ranking strategy for constructing preference lists $\mathcal{T}_{c,l}$ affects intensity learning, we compare three ordering schemes on the same candidate set.
Specifically, we evaluate the proposed rule-based strategy against two acoustic-driven alternatives, where candidates are ranked by their distance to the target speech in terms of average pitch or energy.
As shown in Figure~\ref{fig:analysis} (b), the rule-based ranking achieves the best overall performance among the three strategies.
In comparison, pitch-based ranking performs the worst, while energy-based ranking yields intermediate results.
These findings suggest that rankings derived from a single acoustic attribute provide weaker supervision signals, whereas the rule-based strategy offers a more effective and stable ordering for listwise preference optimization.

\paragraph{Effect of Weighting Mechanism in Listwise Optimization.}
We study the effect of the weighting mechanism $\lambda$ in listwise optimization under identical training settings.
By default, Emo-LiPO adopts an index-based preference labeling scheme for $\psi_{c,l}$, from which the distance-aware weighting term $\lambda_{i,j}$ is derived.
We compare this default setting with two variants:
(i) \textbf{fixed-$\lambda$}, which disables weighting by setting $\lambda_{i,j}=1$ for all pairs; and
(ii) \textbf{score-$\lambda$}, where $\lambda_{i,j}$ is computed from pre-assigned preference score gaps.
As shown in Figure~\ref{fig:analysis} (c), introducing distance-aware weighting consistently improves performance over the fixed-$\lambda$ variant.
Index-based and score-based weighting achieve similar results, indicating that the performance gains mainly stem from the weighting mechanism itself rather than the specific definition of $\psi_{c,l}$.

\paragraph{Impact of Negative Sampling Strategies.}
We study how the sampling strategy for the single negative sample $S_{\bar{c}}$ affects intensity learning in listwise optimization.
By default, our method samples the negative uniformly at random across all intensity levels (\textit{Rand}).
Each preference list includes one negative sample drawn from a non-target emotion category.
We further compare this default strategy with three alternatives that restrict the negative sample to a specific intensity level (\textit{Low}, \textit{Mid}, or \textit{High}).
As shown in Figure~\ref{fig:analysis} (d), the default random strategy achieves the best overall performance, while constraining negatives to a single intensity level consistently degrades performance.
These results indicate that sampling negatives across diverse intensity levels provides more balanced supervision, whereas single-level sampling introduces bias and weakens stable intensity ordering.

%% file: section/06-conclusion.tex
\section{Conclusion}
This work studies fine-grained emotion intensity control in prompt-conditioned LLM-based text-to-speech systems and identifies the limitations of existing supervised and preference-aligned methods in modeling perceptually meaningful intensity variations.
To address this challenge, we propose \textbf{Emo-LiPO}, a listwise preference optimization framework that formulates emotion intensity control as a learning-to-rank problem and explicitly models global intensity ordering under fixed transcripts.
We further introduce \textbf{ESD-plus}, a multi-speaker emotional speech dataset with explicit intensity variations to support systematic training and evaluation.
Extensive automatic and human evaluations demonstrate that Emo-LiPO consistently outperforms strong supervised and DPO-based LLM TTS baselines in emotion accuracy and intensity controllability, while maintaining high speech quality.
Overall, this work highlights the importance of listwise preference modeling for fine-grained emotional control and provides a principled direction for expressive and controllable speech generation.